\documentclass[aps,prb,twocolumn,showpacs]{revtex4-1}
\usepackage{graphicx}
\usepackage{dcolumn}
\usepackage{bm}
\usepackage{amsmath}
\DeclareGraphicsExtensions{.pdf,.eps,.png,.jpg,.mps} 

\begin{document}

\title{Bound states in the continuum: localization of Dirac-like fermions}

\author{N. Cort\'es}
\affiliation{Departamento de F\'{\i}sica, Universidad Cat\'olica
del Norte, Casilla Postal 1280, Antofagasta, Chile}

\author{Leonor Chico}
\affiliation{Instituto de Ciencia de Materiales de Madrid, Consejo Superior de Investigaciones Cient\'{i}ficas (ICMM-CSIC), 
C/ Sor Juana In\'es de la Cruz 3, 28049 Madrid, Spain}

\author{M. Pacheco}
\author{L. Rosales}
\author{P. A. Orellana}
\email{pedro.orellana@usm.cl}
\affiliation{Departamento de F\'{i}sica, Universidad T\'{e}cnica
Federico Santa Mar\'{i}a, Casilla 110-V, Valpara\'{i}so, Chile}

\date{\today}

\begin{abstract}
We report 
the formation of bound states in the continuum for  Dirac-like fermions in 
structures composed by a trilayer graphene flake connected to nanoribbon leads. The
existence of this kind of localized states 
 can be proved by combining 
local density of
states and electronic conductance calculations. By applying a gate voltage, the bound states couple to
the continuum, yielding a maximum in the electronic transmission. This feature can be 
exploited to identify 
bound states in the continuum in graphene-based structures.
\end{abstract}


\maketitle

\section{Introduction}

The electronic behavior of graphene has been the focus of a great amount of work since its isolation in 2004 \cite{Novoselov_2004}. 
Graphene has an outstanding mobility at room temperature, being an excellent metallic material at the nanoscale. 
Since its charge carriers behave as chiral massless Dirac fermions, it is not possible to confine them by purely electrostatic means, 
so other strategies, such as employing bilayer graphene \cite{bilayerzhang}, have to be considered for nanoelectronic applications, for which gapped materials
are needed to fabricate transistors and logic gates. 

The difficulty to confine charge carriers in graphene has a fundamental origin: Klein tunneling was predicted to occur in monolayer graphene \cite{katsnelsonkl}, and it was subsequently 
detected in transport experiments \cite{Young_2009,Stander_2009}. Devising alternative ways to confine massless fermions is therefore a relevant topic in graphene research. 
One possible way to attain localized states in graphene is by means of bound states in the continuum (BICs). 
The 
remarkable 
advances in the fabrication of graphene nanoribbons and 
experiments in which their electronic properties are tuned by applied external potentials,
suggest that BICs could be observed in graphene structures.

Bound states in the continuum
 were first predicted by von Neumann and Wigner in 1929,
who constructed a potential that yielded a truly localized, square-integrable state completely embedded
in a continuum \cite{boundstate1}. Since then, a great number of theoretical works have explored the
feasibility of BICs in diverse setups.
In fact, they have been predicted to occur in atomic and molecular systems \cite{stillinger,friedrich1,cederbaum}, as well as
in mesoscopic structures \cite{schult,zhen-li}.
Systems with quantum dots can be exploited to produce BICs in
low-dimensional structures \cite{olendski,rotter,orellanapssa,loreto,epljhon}.

The formation of BICs is a result of the
interference of resonant states via the continuum. 
Several mechanisms can give rise to these bound states: 
They can appear due to symmetry effects \cite{texier,epljhon}, 
so the difference of parity between discrete states
 and the continuum prevents their coupling. 
Another mechanism takes into account a nonzero coupling between quasi-bound states 
and the continuum, 
 so BICs might be the result of a destructive interference process of resonant states for certain values of the physical parameters of the
 system
 \cite{Miyamoto, Sadreev1, Sadreev2}. Finally, a third mechanism is based in the Fabry-Perot interferometer, of application to photonic systems \cite{Sadreev3}.

As the physics of BICs relies on interference, it is not limited to quantum systems. In fact,  in Ref. \onlinecite{nockel} ballistic transport
through a quantum dot  was studied, demonstrating the possibility of a 
classical analogue of BICs.  
Furthermore, exploiting the analogy between electronics and photonics, several proposals have 
predicted BICs in photonic materials \cite{marinica,evgeny,prodanovic}. And time-dependent fields can be employed to achieve such localized states, as it has been
recently proposed \cite{eplclara2}.

There is only one experimental work reporting the measurement of BICs; remarkably, it has been achieved in a photonic system \cite{plotnik}. 
The observation by Capasso {\emph et al.} \cite{capasso} of an electronic state above the barrier 
of  a semiconductor heterostructure was considered by other authors to be a BIC, but in fact it turned out to lie in a minigap of the 
superlattice \cite{plotnik}. Thus, for the time being, no observation of a bound state in the continuum has been achieved for
electronic systems. Indeed, BICs are considered to be fragile states, difficult to construct and detect experimentally \cite{repdelonghi,zhangprl}.
Therefore, the search of electronic systems which could reveal the existence of robust BICs with unambiguous features 
 is an important field of research. As discussed above, one possibility for detection of BIC is to employ graphene-based devices. 
  
   In this work we show that BICs can be detected in graphene trilayer systems. 
Our findings indicate that this nanostructure is an excellent candidate to
observe bound states in the continuum by combining the measurement of the LDOS and the conductance through the system as a function of the 
gate voltage. This gate potential produces the coupling of the BIC to the continuum, so the LDOS at the BIC energy will be reduced for increasing values of the gate potential; however, we find that this coupling yields a maximum of the transmission through the system that persists for large gate voltages. Thus, the combination of LDOS and transport measurements provides a way to identify BICs in electronic systems.

 In Section II of the paper we describe  the tight-binding model employed to calculate the electronic properties of the trilayer graphene structure.  In Section III we show an analytical approach to find the transmission probability by modeling the trilayer as three coupled 1D chains.  In Section IV we present results for the LDOS and the conductance as a function of a gate voltage and discuss the formation of bound states in the continuum.  Finally, Section V summarizes our main results.

\section{  Model and System}
 \label{sec:modsys}
We employ a  $\pi$-orbital tight-binding model, which gives an excellent description of the electronic properties of graphene systems around the Fermi level.  The in-plane nearest-neighbor interaction given by a single hopping parameter $\gamma$, which we take as $-3$ eV. One interlayer hopping parameter between atoms directly placed on top of each other couples the flakes, being  $\gamma^{\prime}=0.1 \gamma$.

\begin{figure}
\centerline{\includegraphics[width=90mm,clip]{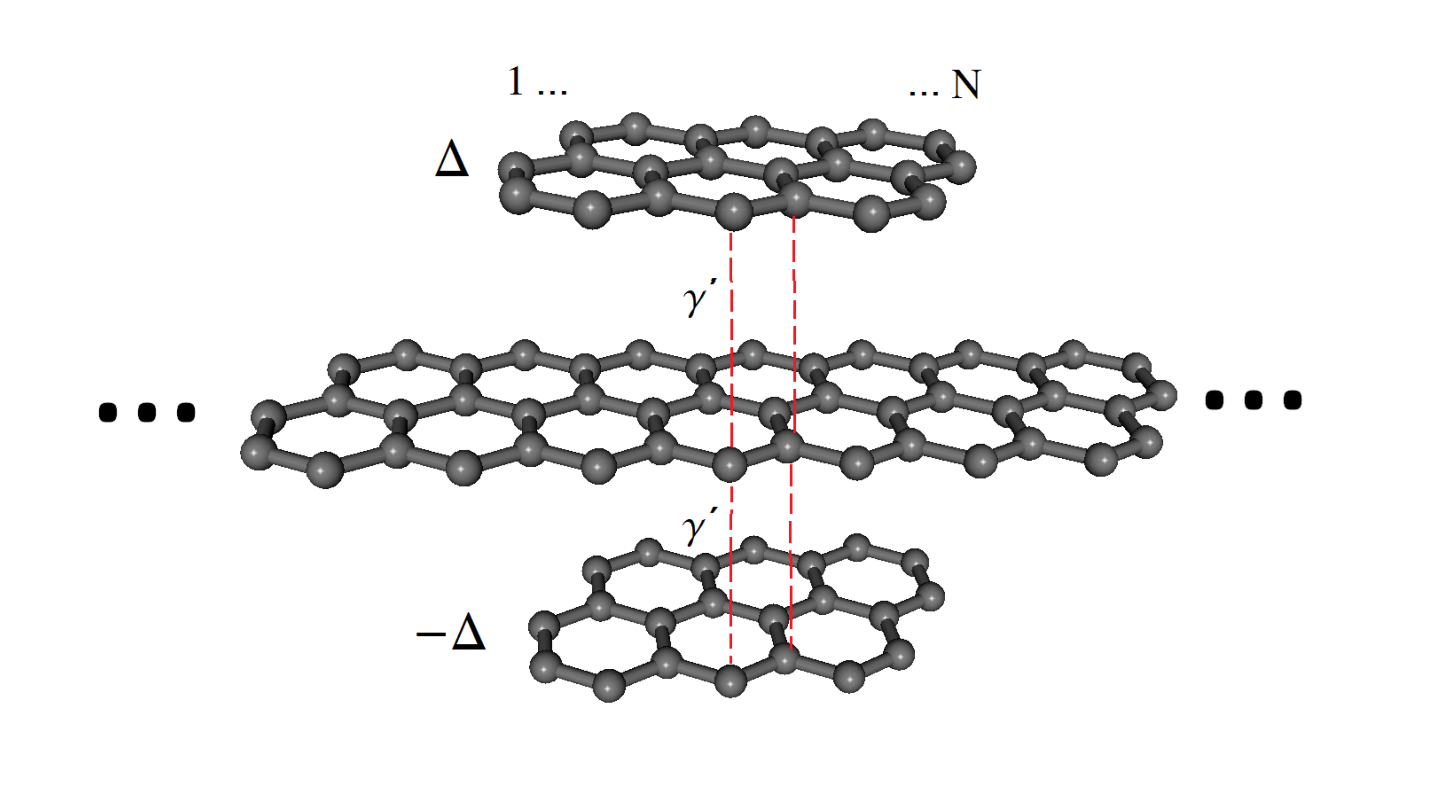}} 
  \caption{Trilayer graphene nanostructure with $AAA$ stacking. $\Delta$ is the gate voltage.}
  \label{fig1}
\end{figure}

We consider a trilayer armchair flake with direct AAA stacking. The leads are also armchair nanoribbons connected to the central flake, as shown in Fig. \ref{fig1}. 
This structure can be alternatively viewed as an infinite armchair nanoribbon with two flakes placed symmetrically above and below it. 
Note that, although Bernal (AB) stacking is more stable for graphite, the direct or AA stacking has been experimentally found in few-layer graphene \cite{Liu_2009}. 
We choose the direct stacking because  the AAA-stacked system can be easily mapped into a one-dimensional chain with two hoppings, yielding analytical expressions of the transmission that perfectly fit the conductance obtained numerically {\it in the first transmission channel}, as we later show in this paper. 
The widths considered correspond to metallic armchair nanoribbons, which play the role of contacts in this system. We give the length of the finite flake in translational unit cells, which corresponds to the number of 4-atom units along it. For the width we use the standard notation, giving the number of dimer chains across the ribbon.

 The total Hamiltonian is described by the sum of each graphene flake (up, central and down) Hamiltonian $H_{f}^{\alpha}$, with $(\alpha=u, c$ and $d)$, the interaction between the flakes $\beta$, $(\beta=u$ and $d$) with the central flake $H_{\beta - {\it cf}}$, the interaction between central flake with the contacts $H_{\it cf-leads}$ and the Hamiltonian of the contacts $H_{\it leads}$:

\begin{equation}\label{Htotal}
H=\sum_{\alpha=u,c,d}H_{f}^{\alpha}+\sum_{\beta=u,d} H_{\beta - {\it cf}}+H_{\it cf - leads}+H_{\it leads},
\end{equation}
with 

\begin{eqnarray}
H_{f}^{\alpha}&=&\sum_{i =1}^{N}\varepsilon_{i}^{\alpha}c_{i}^{\alpha {\dag}}c_{i}^{\alpha}+\gamma\sum_{\substack{i,j=1 \\ \langle i,j \rangle}}^{N}(c_{i}^{\alpha {\dag}}c_{j}^{\alpha}+H.c.),\\
H_{\beta - {\it cf}}&=&\gamma^{\prime}\sum_{i=1}^N (c_{i}^{\beta {\dag}}c_{j}^{c}+H.c.),\\
H_{\it cf -leads}&=&\gamma\left\{(c_{1}^{c {\dag}}c_{0}^{c}+c_{N}^{c {\dag}}c_{N+1}^{c})+H.c.\right\},\\
H_{\it leads}&=&\sum_{\substack{i>N, \\ i<1}}\varepsilon_{i}^{c}c_{i}^{c {\dag}}c_{i}^{c}+\gamma\sum_{\substack{i,j>N,\\ i,j<1\\ \langle i,j \rangle}}(c_{i}^{c {\dag}}c_{j}^{c}+H.c.),
\end{eqnarray}

\noindent where $\varepsilon_{i}^{\alpha}$  is the site energy for atom $i$ in layer $\alpha$;  $c_{i}^{\alpha}$ $(c_{i}^{\alpha {\dag}})$ is the annihilation (creation)
operator of one electron in atom $i$  of layer $\alpha$; $\gamma$ is the nearest-neighbor hopping between atoms inside a layer and
$\gamma^{\prime}$ is the hopping between atoms directly on top of each other in neighboring layers. Note that as the leads are smoothly connected to the central flake constituting a nanoribbon, we leave the superindex $c$ for lead variables and operators.

\section{Analytical solution}
 \label{sec:ansol}

This problem is often solved numerically by standard Green function techniques, but it also admits an analytical approach. To this end, a graphene monolayer is mapped onto a one-dimensional (1D) chain \cite{JAP2013}, so the trilayer is modeled as three coupled  
1D chains.

The stationary state for one monolayer graphene can be written as 
$|\psi\rangle=\sum_{j,m}^{N}\left(\varphi_{j,m}^{A}|j,m\rangle^{A}+\varphi_{j,m}^{B}|j,m\rangle^{B}\right),$
where $\varphi_{j,m}^{A}$ and $\varphi_{j,m}^{B}$ represent the probability amplitudes to find one electron in the dimer $j, m$ in atoms A and B respectively.
From the eigenvalue equation for one monolayer graphene, $H_{layer}^{\alpha}|\psi\rangle=E|\psi\rangle$, two linear difference equations are obtained \cite{JAP2013}. Considering plane wave solutions with $q$ transverse wavenumber,  $\varphi_{j,m}^{A,B}=e^{iqm}\phi_{j}^{A,B}$, the equations of motion are

\begin{eqnarray}\label{ecsmovb}
E\phi_{j}^{A}=\gamma(2\cos{q}\hspace{1mm}\phi_{j-1}^{B}+\phi_{j}^{B}),\\
E\phi_{j}^{B}=\gamma(2\cos{q}\hspace{1mm}\phi_{j+1}^{A}+\phi_{j}^{A}).\nonumber
\end{eqnarray}

This is equivalent to a one-dimensional chain with different hoppings between A-B and B-A atoms, namely, $\gamma$ and $\eta=2 \gamma \cos q$. 
Incidentally, note that close to the Dirac point $K$, $q$ takes the value  $2\pi /3$,  so the equations are written as 

\begin{eqnarray}\label{ecsmovb2}
E\phi_{j}^{A}=\gamma(-\hspace{1mm}\phi_{j-1}^{B}+\phi_{j}^{B}),\\
E\phi_{j}^{B}=\gamma(-\hspace{1mm}\phi_{j+1}^{A}+\phi_{j}^{A}).\nonumber
\end{eqnarray}
Thus, from the tight-binding equations of motion near the $K$ point, we arrive in the continuum limit at the Dirac equation, 
$E\tilde{\Phi}(x) = v_F \sigma_2\partial_x\tilde{\Phi}(x)$, where $v_F=\gamma a/\hbar$ is the Fermi velocity and $a$ is the lattice constant of graphene.

\begin{figure}
\includegraphics[width=85mm,clip]{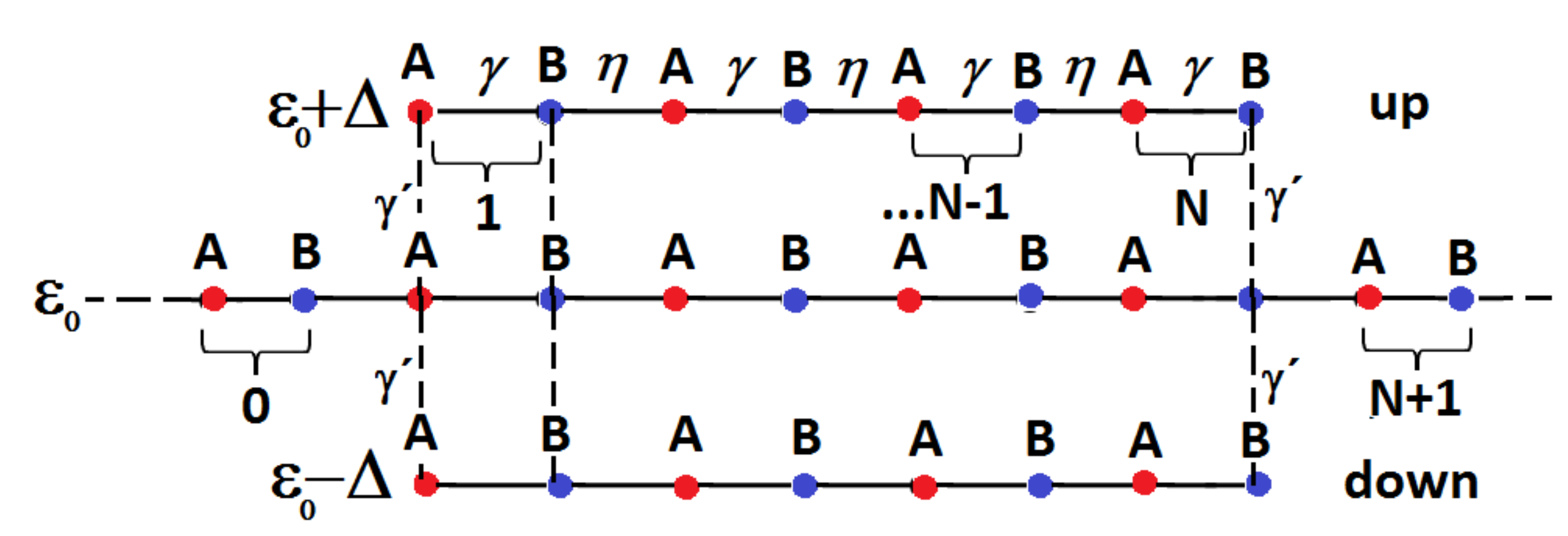}
\caption{Trilayer graphene flake connected to two semiinfinite ribbons represented as three diatomic chains with different hoppings.}
\label{trilayer}
\end{figure}

\subsection{Decoupling trilayer graphene modes}\label{autovalyautovec}
Now we consider the trilayer system as composed of three 1D chains with interchain hopping $\gamma^{\prime}$, as depicted in Fig. \ref{trilayer}. The equations of motion for trilayer graphene can be written in a
basis that decouples the transversal modes.
The Hamiltonian representing this system is given by:
\begin{equation}\label{hamiltoniano}
H=    \left(
      \begin{array}{ccc}
        \varepsilon_{\rm 1D}+\Delta & \gamma\hspace{0.01cm}^{\prime}  & 0 \\
        \gamma\hspace{0.01cm}^{\prime} & \varepsilon_{\rm 1D} & \gamma\hspace{0.01cm}^{\prime} \\
        0 & \gamma\hspace{0.01cm}^{\prime} &  \varepsilon_{\rm 1D}-\Delta \\
      \end{array}
    \right), 
\end{equation}
Where $\varepsilon_{\rm 1D}$ is the energy of the one-dimensional chain, and $\Delta$ is the gate potential. 
The above Hamiltonian is written in the basis  
\begin{equation}\label{base}
\Psi=\left(
  \begin{array}{c}
    \Psi^{u} \\
    \Psi^{c} \\
    \Psi^{d} \\
  \end{array}
\right),
\end{equation}
where $\Psi^{u}, \Psi^{c}$ and  $\Psi^{d}$ represent the atomic wavefunctions of the chain \emph{up}, \emph{central} and \emph{down} respectively. 

Solving  the eigenvalue equation  $H|\Psi\rangle=\varepsilon|\Psi\rangle$ we obtain
\begin{eqnarray}\label{autovalores}
\nonumber\overline{\varepsilon}_{1}&=&\varepsilon_{\rm 1D}+\sqrt{\Delta^{2}+2\gamma\hspace{0.01cm}^{\prime^{2}}}, \\
         \overline{\varepsilon}_{2}&=&\varepsilon_{\rm 1D}, \\
\nonumber\overline{\varepsilon}_{3}&=&\varepsilon_{\rm 1D}-\sqrt{\Delta^{2}+2\gamma\hspace{0.01cm}^{\prime^{2}}}.
\end{eqnarray}
The energies  $\overline{\varepsilon}_{1}, \overline{\varepsilon}_{2}$ y $\overline{\varepsilon}_{3}$ are the renormalized energies for the 
atoms of the uncoupled chains \emph{up}, \emph{central} and \emph{down} respectively.  The corresponding eigenvectors form a new basis which diagonalizes the Hamiltonian $\bar{H}=P^{-1}HP$, and therefore decouples the three chains, where  the matrix $P$ is 
\begin{equation}\label{matrizP}
P=    \left(
      \begin{array}{ccc}
        \frac{\gamma\hspace{0.01cm}^{\prime}}{\sqrt{\Delta^{2}+2\gamma\hspace{0.01cm}^{\prime^{2}}}-\Delta} & 1 & -\frac{\gamma\hspace{0.01cm}^{\prime}}{\sqrt{\Delta^{2}+2\gamma\hspace{0.01cm}^{\prime^{2}}}+\Delta} \\
        1 & -\frac{\Delta}{\gamma\hspace{0.01cm}^{\prime}} & 1 \\
        \frac{\gamma\hspace{0.01cm}^{\prime}}{\sqrt{\Delta^{2}+2\gamma\hspace{0.01cm}^{\prime^{2}}}+\Delta} & -1 &  \frac{\gamma\hspace{0.01cm}^{\prime}}{-\sqrt{\Delta^{2}+2\gamma\hspace{0.01cm}^{\prime^{2}}}+\Delta}
      \end{array}
    \right),
\end{equation}
so that 
\begin{equation}\label{jjj}
\bar{H}=\left(
          \begin{array}{ccc}
            \overline{\varepsilon}_{1} & 0 & 0 \\
            0 & \overline{\varepsilon}_{2} & 0 \\
            0 & 0 & \overline{\varepsilon}_{3} \\
          \end{array}
        \right).
\end{equation}

\subsection{Transversal modes of the trilayer system}

We can write the equations of motion 
for sites $j=1$ and  $j=N$ of the trilayer flake depicted in Fig. \ref{trilayer}. By representing these equations in the basis of eigenvectors of matrix $P$,  the new equations of motion for the chains are  decoupled in the transversal normal modes $u_{j}, v_{j}$ and $w_{j}$ for the renormalized chains, given by 

\begin{eqnarray}\label{modnorm}
\nonumber u_{j}&=&\frac{\gamma\hspace{0.01cm}^{\prime}}{\sqrt{\Delta^{2}+2\gamma\hspace{0.01cm}^{\prime^{2}}}-\Delta}\Psi_{j}^{u}+\Psi_{j}^{c}+\frac{\gamma\hspace{0.01cm}^{\prime}}{\sqrt{\Delta^{2}+2\gamma\hspace{0.01cm}^{\prime^{2}}}+\Delta}\Psi_{j}^{d},\\
v_{j}&=&\Psi_{j}^{u}-(\Delta/\gamma\hspace{0.01cm}^{\prime})\Psi_{j}^{c}-\Psi_{j}^{d},\\
\nonumber w_{j}&=&\frac{-\gamma\hspace{0.01cm}^{\prime}}{\sqrt{\Delta^{2}+2\gamma\hspace{0.01cm}^{\prime^{2}}}+\Delta}\Psi_{j}^{u}+\Psi_{j}^{c}-\frac{\gamma\hspace{0.01cm}^{\prime}}{\sqrt{\Delta^{2}+2\gamma\hspace{0.01cm}^{\prime^{2}}}-\Delta}\Psi_{j}^{d}.
\end{eqnarray}
For site $j=1$ we have

\begin{eqnarray}\label{ecsmovj1}
\nonumber  (\varepsilon_{\rm 1D}-\overline{\varepsilon}_{1})u_{1}&=&u_{2}+\Psi_{0}^{c}, \\
  (\varepsilon_{\rm 1D}-\overline{\varepsilon}_{2})v_{1}&=&v_{2}-(\Delta/\gamma\hspace{0.01cm}^{\prime})\Psi_{0}^{c}, \\
\nonumber  (\varepsilon_{\rm 1D}-\overline{\varepsilon}_{3})w_{1}&=&w_{2}+\Psi_{0}^{c},
\end{eqnarray}
and for site  $j=N$, 
\begin{eqnarray}\label{ecsmovjn}
\nonumber  (\varepsilon_{\rm 1D}-\overline{\varepsilon}_{1})u_{N}&=&u_{N-1}+\Psi_{N+1}^{c}, \\
(\varepsilon_{\rm 1D}-\overline{\varepsilon}_{2})v_{N}&=&v_{N-1}-(\Delta/\gamma\hspace{0.01cm}^{\prime})\Psi_{N+1}^{c}, \\
\nonumber  (\varepsilon_{\rm 1D}-\overline{\varepsilon}_{3})w_{N}&=&w_{N-1}+\Psi_{N+1}^{c},
\end{eqnarray}
where $\Psi_{0}^{c}$ y $\Psi_{N+1}^{c}$  are the electron wavefunctions in the central chain for sites  $j=0$ and $j=N+1$ respectively. 

To obtain the probability amplitude of each renormalized chain, we consider the scattering solution in the three regions of the problem, namely, the left 
lead/ribbon, 
\begin{equation}\label{zona1}
\Psi_{j}^{\text{I}}=e^{ik_{\rm 1D}j}+re^{-ik_{\rm 1D}j}\quad,\quad\,\,\,-\infty\leq\Psi_{j}^{\text{I}}\leq0,
\end{equation}
the central trilayer, 
\begin{equation}\label{modosnormales}
 \left\lbrace
\begin{array}{lll}
u_{j}=A_{1}e^{iq_{1}j}+B_{1}e^{-iq_{1}j}\\
v_{j}=A_{2}e^{iq_{2}j}+B_{2}e^{-iq_{2}j}\quad, 0\leq u_{j},v_{j}\hspace{0.1cm}\text{and 
}\hspace{0.1cm}w_{j}\leq N+1,\\
w_{j}=A_{3}e^{iq_{3}j}+B_{3}e^{-iq_{3}j}
\end{array}
\right.
\end{equation}
and the right lead/ribbon, 
\begin{equation}\label{zona3}
\Psi_{j}^{\text{III}}=te^{ik_{\rm 1D}j}\quad,\quad \hspace{1mm} N+1\leq\Psi_{j}^\text{{III}}\leq +\infty,
\end{equation}\\
where  $A_{i}$ and $B_{i}$,  ($i=1,2$ and $3$) are the probability amplitudes of the three renormalized chains. As it is customary, $r$ and $t$ are the reflection and transmission coefficients. Notice that $q_{i}$  ($i=1,2$ and $3$) are the allowed wave numbers  in each renormalized chain, given in Eq. 9 of the paper. 

By replacing the expressions for the normal modes given by Eq. \ref{modosnormales} in the new equations of motion Eqs. \ref{ecsmovj1} and \ref{ecsmovjn} for sites $j=1$ and $j=N$, and by taking into account  the expressions for the scattering  wavefunctions at the left and right leads/regions, given by Eqs. \ref{zona1} and \ref{zona3}, we obtain a system of six equations, 

\begin{eqnarray}\label{sixeqn}
\nonumber A_{1}+B_{1}&=&(r+1),\\
\nonumber A_{1}e^{iq_{1}(N+1)}+B_{1}e^{-iq_{1}(N+1)}&=&te^{ik_{\rm 1D}(N+1)},\\
A_{2}+B_{2}&=&-\frac{\Delta}{\gamma\hspace{0.01cm}^{\prime}}(r+1),\\
\nonumber A_{2}e^{iq_{2}(N+1)}+B_{2}e^{-iq_{2}(N+1)}&=&-\frac{\Delta}{\gamma\hspace{0.01cm}^{\prime}}te^{ik_{\rm 1D}(N+1)},\\
\nonumber A_{3}+B_{3}&=&(r+1),\\
\nonumber A_{3}e^{iq_{3}(N+1)}+B_{3}e^{-iq_{3}(N+1)}&=&te^{ik_{\rm 1D}(N+1)},
\end{eqnarray}
Applying the boundary conditions to the wavefunctions of the flake and leads (in sites $j=0$ and $j=N+1$), we obtain two additional equations, 

\begin{eqnarray}\label{twoeqn}
\nonumber\frac{\gamma\hspace{0.01cm}^{\prime}\Delta}{\Delta^{2}+2\gamma\hspace{0.01cm}^{\prime^{2}}} &&\left[ \frac{\gamma\hspace{0.01cm}^{\prime}}{\Delta}(u_{1}+w_{1})-v_{1}\right]=re^{-ik_{\rm 1D}}+e^{ik_{\rm 1D}}\\
 \frac{\gamma\hspace{0.01cm}^{\prime}\Delta}{\Delta^{2}+2\gamma\hspace{0.01cm}^{\prime^{2}}} &&\left[ \frac{\gamma\hspace{0.01cm}^{\prime}}{\Delta}(u_{N}+w_{N})-v_{N}\right]=te^{ik_{\rm 1D}N}.
\end{eqnarray}
Solving the system of six equations \ref{sixeqn},  we obtain the probability amplitudes $A_{i}$ and  $B_{i}$, which substituted in turn into Eqs.  \ref{modosnormales} 
yields for $j=1$ to the following expressions, 

\begin{eqnarray}\label{modnorm1}
\nonumber u_{1}&=&(r+1)\frac{U_{N-1}(q_{1})}{U_{N}(q_{1})}+\frac{te^{ik_{\rm 1D}(N+1)}}{U_{N}(q_{1})},\\
v_{1}&=&\nu\left[(r+1)\frac{U_{N-1}(q_{2})}{U_{N}(q_{2})}+\frac{te^{ik_{\rm 1D}(N+1)}}{U_{N}(q_{2})}\right],\\
\nonumber w_{1}&=&(r+1)\frac{U_{N-1}(q_{3})}{U_{N}(q_{3})}+\frac{te^{ik_{\rm 1D}(N+1)}}{U_{N}(q_{3})},
\end{eqnarray}
and for $j=N$, 
\begin{eqnarray}\label{modnormN}
\nonumber u_{N}&=&\frac{(r+1)}{U_{N}(q_{1})}+te^{ik_{\rm 1D}(N+1)}\frac{U_{N-1}(q_{1})}{U_{N}(q_{1})},\\
v_{N}&=&\nu\left[\frac{(r+1)}{U_{N}(q_{2})}+te^{ik_{\rm 1D}(N+1)}\frac{U_{N-1}(q_{2})}{U_{N}(q_{2})}\right],\\
\nonumber w_{N}&=&\frac{(r+1)}{U_{N}(q_{3})}+te^{ik_{\rm 1D}(N+1)}\frac{U_{N-1}(q_{3})}{U_{N}(q_{3})}.
\end{eqnarray}
Here $\nu=-\Delta/\gamma^{\prime}$ and $U_n(x)$ are the Chebyshev polynomials of the second kind, where $x=q_1,q_2, q_3$ are the allowed wavevectors in the renormalized chains composing the trilayer,
\begin{eqnarray}
\nonumber q_{1}&=&\arccos{\left[\frac{\varepsilon_{\rm 1D}-\sqrt{\Delta^{2}+2\gamma^{\prime^{2}}}}{2}\right]},\\
q_{2}&=&\arccos\left[\frac{\varepsilon_{\rm 1D}}{2}\right],\\
\nonumber q_{3}&=&\arccos{\left[\frac{\varepsilon_{\rm 1D}+\sqrt{\Delta^{2}+2\gamma^{\prime^{2}}}}{2}\right]}.  
\end{eqnarray}
 Notice that these three wavevectors correspond to the three states of the trilayer, namely, the coupled bonding and antibonding solutions and the nonbonding state, which is independent of the interlayer coupling. This latter state gives rise to the BIC. 
We substitute  Eqs. \ref{modnorm1} and  \ref{modnormN} in Eq. \ref{twoeqn}, so we obtain a system of two equations for the reflection and transmission coefficients $r$ and $t$,

\begin{eqnarray}\label{ecsryt}
e^{ik_{\rm 1D}}+re^{-ik_{\rm 1D}}&=&\mu[(r+1)G+te^{ik_{\rm 1D}(N+1)}F],\\
\nonumber te^{ik_{\rm 1D}N}&=&\mu[(r+1)F+te^{ik_{\rm 1D}(N+1)}G],
\end{eqnarray}
here   $\mu=\gamma^{\prime}/(\Delta^{2}+2\gamma^{\prime^{2}})$, $k_{\rm 1D}$ is the wavevector of the 1D chain and $H=F^2-G^2$ with   $F$, $G$  given by 

 \begin{eqnarray}
\nonumber  F &= & \frac{\gamma^{\prime}}{U_{N}(q_{1})}-\frac{\Delta\nu}{U_{N}(q_{2})}+\frac{\gamma^{\prime}}{U_{N}(q_{3})},\\
\nonumber  G &= &\gamma^{\prime}\frac{U_{N-1}(q_{1})}{U_{N}(q_{1})}-\Delta\nu\frac{U_{N-1}(q_{2})}{U_{N}(q_{2})}+\gamma^{\prime}\frac{U_{N-1}(q_{3})}{U_{N}(q_{3})},
\end{eqnarray}

 The solution of this system of equations gives an analytical expression for the transmission probability through the trilayer,

\begin{widetext}
\begin{equation}
T=|t|^{2}=\frac{2\mu^{2}F^{2}\left(1-\cos{2k_{\rm 1D}}\right)}{\mu^{4}H^{2}+4\mu^{2}G^{2}-4\mu G\cos{k_{\rm 1D}}+2\mu^{2}H\left(2\mu G\cos{k_{\rm 1D}}-\cos{2k_{\rm 1D}}\right)+1},
 \end{equation}
 \end{widetext}

\section{Results and Discussion}
 \label{sec:results}

 From the theoretical perspective, a BIC can be identified by a Dirac delta-like peak in the LDOS embedded in the continuum, being a square-integrable, localized state coexisting with non-normalizable, propagating states.  However, this does not constitute an unequivocal description from the experimental viewpoint, given that resonances or Fabry-P\'erot states can also give rise to very sharp peaks in the density of states. 
We have studied the evolution of the LDOS and the conductance of these BIC states in graphene trilayers as a function of the gate voltage $\Delta$, showing that they can be identified unambiguously considering both LDOS and transport measurements.

In order to check our results, we also obtain numerically the conductance and the LDOS for these systems employing a decimation technique to obtain the Green's function of the structure, as done previously for similar systems \cite{Jhon_2010,Jhon_2012}. In the energy region with only one propagating state in the leads, the
results agree perfectly. Within the coupled chains model, the LDOS is obtained as the sum of the squared moduli of the amplitudes over the trilayer flake.

\begin{figure}
\includegraphics[width=\columnwidth,clip]{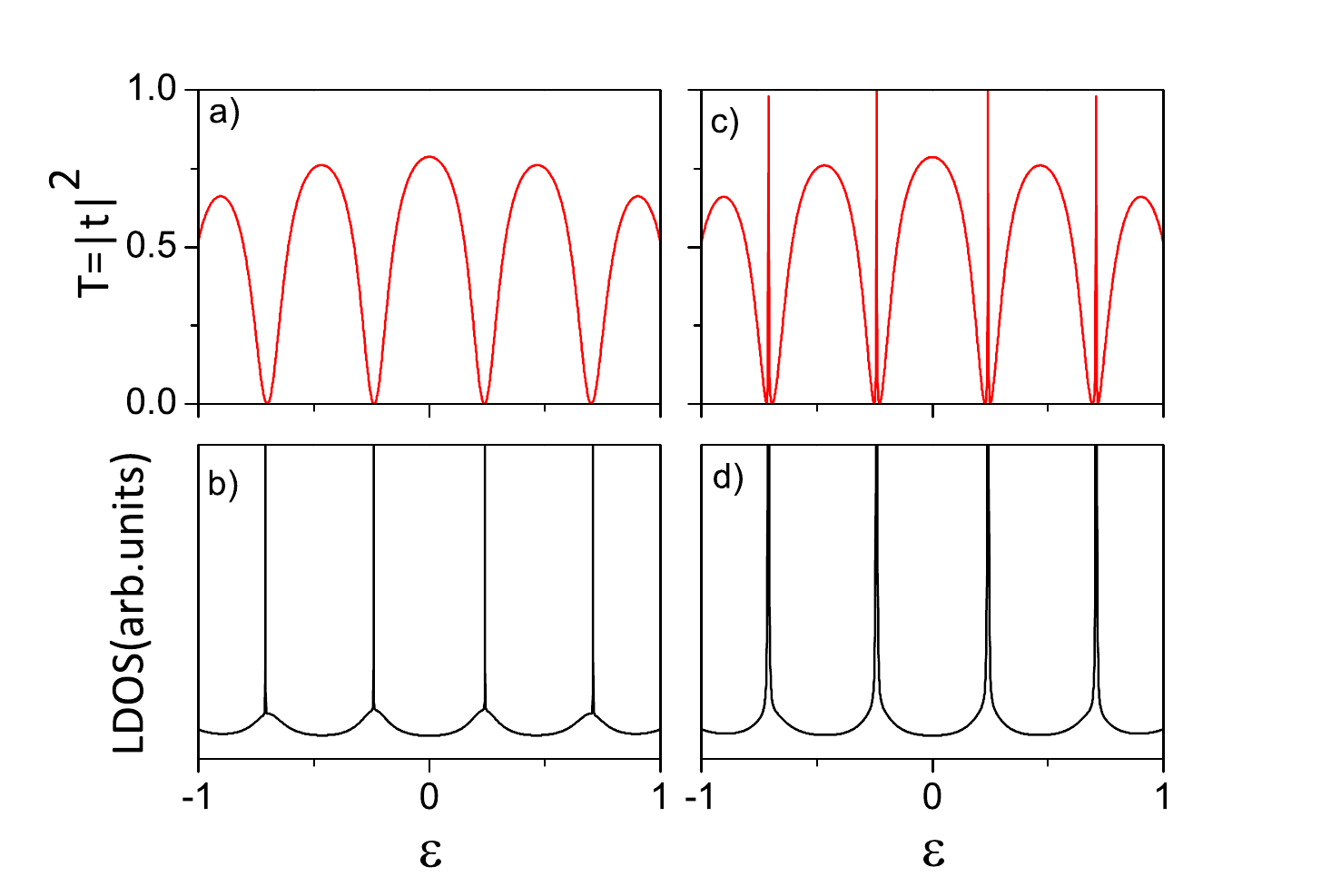}
\caption{Transmission and LDOS versus energy for the armchair trilayer $N=12$ ribbon for $\Delta=0$ (left panels) and $\Delta = 0.01$.
}
\label{tryldos1}
\end{figure}

In Fig. \ref{tryldos1} we present the transmission and LDOS for the $N=12$ trilayer calculated with the coupled chain model for the symmetric case ($\Delta =0$) and for a nonzero, albeit small, value of the gate potential, $\Delta=0.01$. Energies are given in units of $\gamma$ throughout the paper. 
The LDOS shows very sharp peaks superposed to broader peaks over a nonzero density for all energies. The  
sharp features correspond to BICs. The broader peaks correspond to antiresonances related to bonding and antibonding states of the trilayer,  
strongly coupled to the continuum. The shape of these superimposed peaks can be described by the sum of two Lorentzians with quite different line shapes: one  corresponds to the strongly coupled state, and the other to the BIC, yielding a Dirac delta in the limit of zero gate voltage \cite{epljhon}.

The transmission probability, shown in the top panels of Fig.  \ref{tryldos1},  presents antiresonances at the energies of the strongly coupled states, but there is not any signature of BICs in the transport properties for $\Delta =0$.  However, for $\Delta=0.01$ a sharp transmission peak appears for each BIC, reaching the maximum value $T=1$. This change indicates that the BICs are beginning to couple to the continuum, being in fact quasi-bound states in the continuum (quasi-BICs). Due to this 
coupling, they contribute to the conductance of the system. The LDOS still shows the maxima at the energies of the quasi-BICs. 

\begin{figure}
\includegraphics[width=\columnwidth,clip]{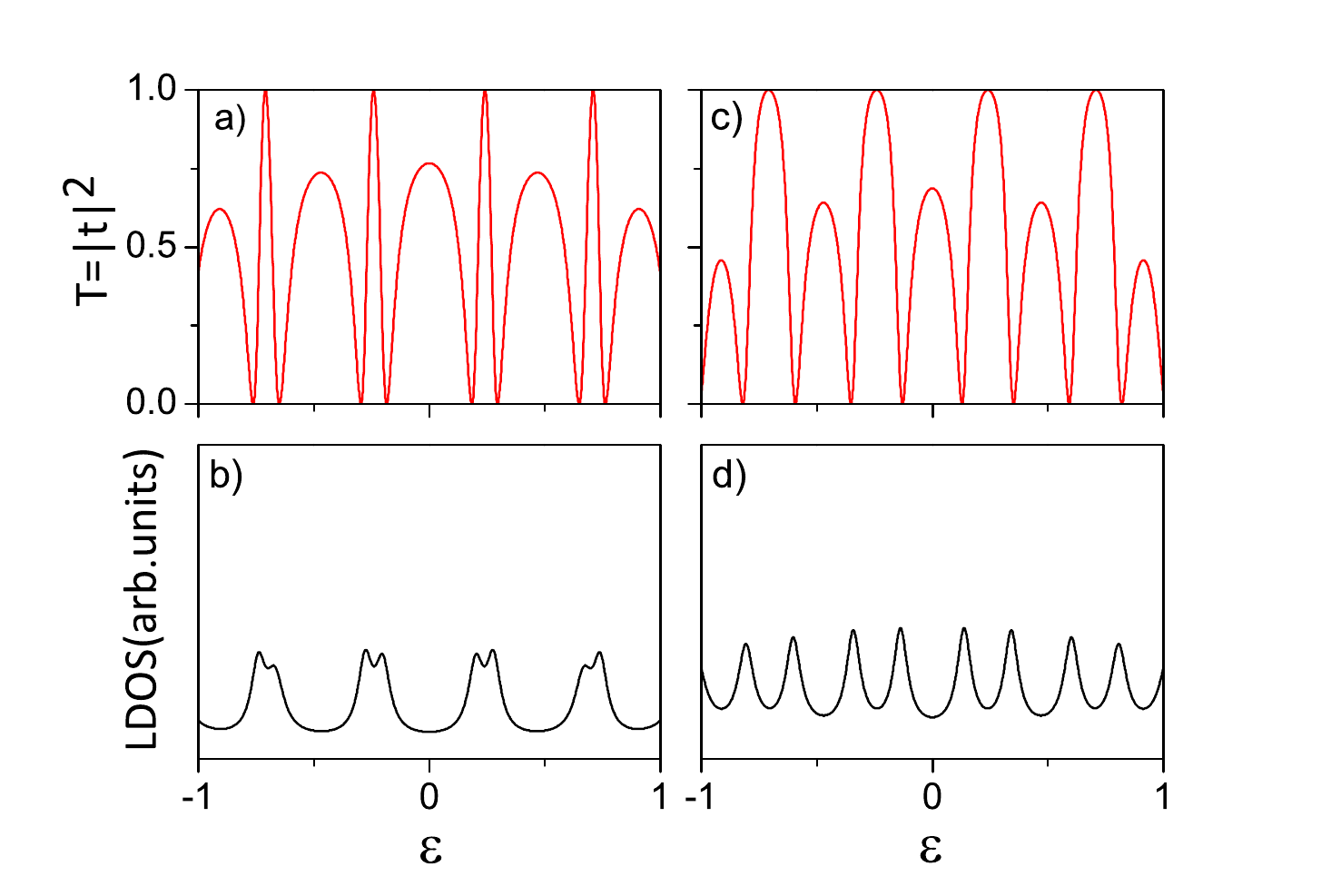}
\caption{Transmission and LDOS versus energy for the armchair trilayer $N=12$ ribbon for $\Delta=0.05$ (left panels) and $\Delta = 0.1$.
}
\label{tryldos2}
\end{figure}

\begin{figure}
\includegraphics[width=\columnwidth,clip]{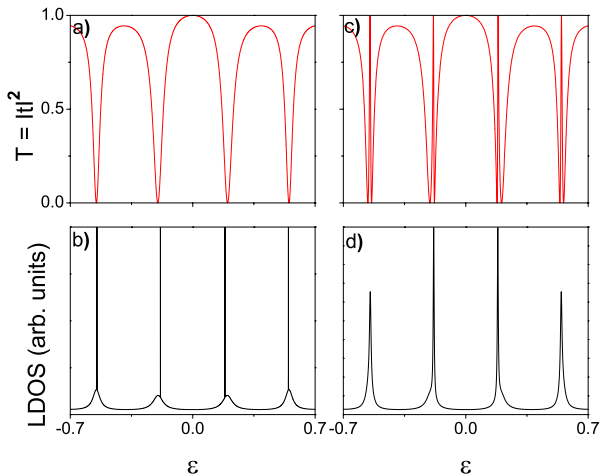}
\caption{Transmission and LDOS versus energy for the armchair trilayer with stacking ABA. Parameters as in Fig. \ref{tryldos1}.
}
\label{ABA}
\end{figure}

In fact, the gate potential plays the role of the asymmetry parameter, which controls the coupling of quasi-BICs to the continuum in this system. Remarkably, increasing the value of the gate potential does not destroy the characteristic maximum in the transmission. Fig. \ref{tryldos2} shows the LDOS and transmission probabilities for larger values of the
gate voltage, $\Delta=0.05$ and $\Delta=0.1$. The coupling of the quasi-BICs to the continuum is so large that they no longer show any peak in the LDOS; however, their presence can be inferred from the maxima in the transmission that persist for these voltages. Most notably, the maximum value of the transmission still reaches 1, while the other maxima appearing between resonances have a smaller value, dependent on the length of the flake, similarly to the results found for bilayer flakes \cite{Jhon_2010,Acta_2012}. 
Besides, the width of the maxima in the transmission corresponding to these quasi-BIC states increases with the value of the gate potential (the coupling parameter), thus yielding a robust signal that can be studied as a proof of the existence of BICs in trilayer graphene systems. 
In order to analyse  the influence of the flakes stacking in our results we have also numerically calculated the LDOS and transmission probability for a ABA-stacked system. Fig. \ref{ABA} shows the transmission probability and LDOS for a system with the same parameters of Fig. \ref{tryldos1}. We have found the same features characteristic of the presence of BICs. This result shows that the formation  of BICs is independently of the stacking.

When $\Delta\rightarrow 0$, the transmission around the energy of the quasi-BICs can be written as a superposition of Fano and Breit-Wigner line-shapes, 

\begin{equation}
T(\varepsilon)=\frac{(\varepsilon-\varepsilon_{0})^2}{(\varepsilon-\varepsilon_{0})^2+\Gamma_{+}^2}+\frac{\Gamma_{-}^2}{(\varepsilon-\varepsilon_{0})^2+\Gamma_{-}^2}
\end{equation}
where $\varepsilon_0$ is the position of the quasi-BIC, $\Gamma_{+}=2\Gamma_0$ and  $\Gamma_{-}=\Delta^2/2\Gamma_0$. The parameter $\Gamma_0$ is related to the interlayer coupling $\gamma^{\prime}$, so that $T=1$ for zero interlayer coupling and it yields a peak of width $\propto \Delta^2$, as it can be appreciated in Figs. \ref{tryldos1} and \ref{tryldos2}. 
This dependence of the width of the transmission peak as a function of the gate voltage is another signature of the quasi-BIC states. 

  Although pure BICs do not 
give any contribution to the conductance when they are uncoupled, 
coupling them to the continuum by means of a gate voltage 
makes them  contribute to the current with a maximum transmission, with a characteristic dependence on the external potential. The persistence of the maximum, even for large values of the applied
voltage, enables their characterization by the analysis of the LDOS and electronic conductance.

\vskip 10pt

\section{Summary}
\label{sec:sum}

 In this work we have  shown that bound states in the continuum can be detected in graphene trilayer systems. We find that this kind of nanostructures are very adequate to
observe these states  by  measurements of the LDOS and the conductance through the system as a function of the 
gate voltage. This gate potential produces the coupling of the BIC to the continuum, so the LDOS at the BIC energy will be reduced for increasing values of the gate potential; however, we find that this coupling yields a maximum of the transmission through the system that persists for large gate voltages. Thus, the combination of LDOS and transport measurements provides a way to identify BICs in electronic systems.

\begin{acknowledgments}
This work has been partially supported by  Chilean FONDECYT grants 1140571 (P.O.), 1140388 (L. R.), and DGIP/USM internal grant 11.14.68 (M.P.), CONICYT ACT 1204 (L.R., M.P., P.O.) and by the Spanish DGES under grant FIS2012-33521.
L. C. gratefully acknowledges the hospitality of the Universidad T\'ecnica Federico Santa Mar\'{\i}a (Chile). 
\end{acknowledgments}


\begin{thebibliography}{35}
\expandafter\ifx\csname natexlab\endcsname\relax\def\natexlab#1{#1}\fi
\expandafter\ifx\csname bibnamefont\endcsname\relax
  \def\bibnamefont#1{#1}\fi
\expandafter\ifx\csname bibfnamefont\endcsname\relax
  \def\bibfnamefont#1{#1}\fi
\expandafter\ifx\csname citenamefont\endcsname\relax
  \def\citenamefont#1{#1}\fi
\expandafter\ifx\csname url\endcsname\relax
  \def\url#1{\texttt{#1}}\fi
\expandafter\ifx\csname urlprefix\endcsname\relax\def\urlprefix{URL }\fi
\providecommand{\bibinfo}[2]{#2}
\providecommand{\eprint}[2][]{\url{#2}}

\bibitem[{\citenamefont{K.S.Novoselov et~al.}(2004)\citenamefont{K.S.Novoselov,
  A.K.Geim, S.V.Mozorov, D.Jiang, Y.Zhang, S.V.Dubonos, I.V.Gregorieva, and
  A.A.Firsov}}]{Novoselov_2004}
\bibinfo{author}{\bibnamefont{K.S.Novoselov}},
  \bibinfo{author}{\bibnamefont{A.K.Geim}},
  \bibinfo{author}{\bibnamefont{S.V.Mozorov}},
  \bibinfo{author}{\bibnamefont{D.Jiang}},
  \bibinfo{author}{\bibnamefont{Y.Zhang}},
  \bibinfo{author}{\bibnamefont{S.V.Dubonos}},
  \bibinfo{author}{\bibnamefont{I.V.Gregorieva}}, \bibnamefont{and}
  \bibinfo{author}{\bibnamefont{A.A.Firsov}}, \bibinfo{journal}{Science}
  \textbf{\bibinfo{volume}{306}}, \bibinfo{pages}{666} (\bibinfo{year}{2004}).

\bibitem[{\citenamefont{Zhang et~al.}(2009)\citenamefont{Zhang, Tang, Girit,
  Hao, Martin, Zettl, Crommie, Shen, and Wang}}]{bilayerzhang}
\bibinfo{author}{\bibfnamefont{Y.}~\bibnamefont{Zhang}},
  \bibinfo{author}{\bibfnamefont{T.-T.} \bibnamefont{Tang}},
  \bibinfo{author}{\bibfnamefont{C.}~\bibnamefont{Girit}},
  \bibinfo{author}{\bibfnamefont{Z.}~\bibnamefont{Hao}},
  \bibinfo{author}{\bibfnamefont{M.~C.} \bibnamefont{Martin}},
  \bibinfo{author}{\bibfnamefont{A.}~\bibnamefont{Zettl}},
  \bibinfo{author}{\bibfnamefont{M.~F.} \bibnamefont{Crommie}},
  \bibinfo{author}{\bibfnamefont{Y.~R.} \bibnamefont{Shen}}, \bibnamefont{and}
  \bibinfo{author}{\bibfnamefont{F.}~\bibnamefont{Wang}},
  \bibinfo{journal}{Nature} \textbf{\bibinfo{volume}{459}},
  \bibinfo{pages}{820} (\bibinfo{year}{2009}).

\bibitem[{\citenamefont{Katsnelson et~al.}(2006)\citenamefont{Katsnelson,
  Novoselov, and Geim}}]{katsnelsonkl}
\bibinfo{author}{\bibfnamefont{M.~I.} \bibnamefont{Katsnelson}},
  \bibinfo{author}{\bibfnamefont{K.~S.} \bibnamefont{Novoselov}},
  \bibnamefont{and} \bibinfo{author}{\bibfnamefont{A.~K.} \bibnamefont{Geim}},
  \bibinfo{journal}{Nat. Phys.} \textbf{\bibinfo{volume}{2}},
  \bibinfo{pages}{620} (\bibinfo{year}{2006}).

\bibitem[{\citenamefont{Young and Kim}(2009)}]{Young_2009}
\bibinfo{author}{\bibfnamefont{A.~F.} \bibnamefont{Young}} \bibnamefont{and}
  \bibinfo{author}{\bibfnamefont{P.}~\bibnamefont{Kim}},
  \bibinfo{journal}{Nature Phys.} \textbf{\bibinfo{volume}{5}},
  \bibinfo{pages}{222} (\bibinfo{year}{2009}).

\bibitem[{\citenamefont{Stander et~al.}(2009)\citenamefont{Stander, Huard, and
  Goldhaber-Gordon}}]{Stander_2009}
\bibinfo{author}{\bibfnamefont{N.}~\bibnamefont{Stander}},
  \bibinfo{author}{\bibfnamefont{B.}~\bibnamefont{Huard}}, \bibnamefont{and}
  \bibinfo{author}{\bibfnamefont{D.}~\bibnamefont{Goldhaber-Gordon}},
  \bibinfo{journal}{Phys. Rev. Lett.} \textbf{\bibinfo{volume}{102}},
  \bibinfo{pages}{026807} (\bibinfo{year}{2009}).

\bibitem[{\citenamefont{von Neumann and Wigner}(1929)}]{boundstate1}
\bibinfo{author}{\bibfnamefont{J.}~\bibnamefont{von Neumann}} \bibnamefont{and}
  \bibinfo{author}{\bibfnamefont{E.}~\bibnamefont{Wigner}},
  \bibinfo{journal}{Phys. Z.} \textbf{\bibinfo{volume}{30}},
  \bibinfo{pages}{465} (\bibinfo{year}{1929}).

\bibitem[{\citenamefont{Stillinger and Herrick}(1975)}]{stillinger}
\bibinfo{author}{\bibfnamefont{F.~H.} \bibnamefont{Stillinger}}
  \bibnamefont{and} \bibinfo{author}{\bibfnamefont{D.~R.}
  \bibnamefont{Herrick}}, \bibinfo{journal}{Phys. Rev. A}
  \textbf{\bibinfo{volume}{11}}, \bibinfo{pages}{446} (\bibinfo{year}{1975}),
  \urlprefix\url{http://link.aps.org/doi/10.1103/PhysRevA.11.446}.

\bibitem[{\citenamefont{Friedrich and Wintgen}(1985)}]{friedrich1}
\bibinfo{author}{\bibfnamefont{H.}~\bibnamefont{Friedrich}} \bibnamefont{and}
  \bibinfo{author}{\bibfnamefont{D.}~\bibnamefont{Wintgen}},
  \bibinfo{journal}{Phys. Rev. A} \textbf{\bibinfo{volume}{31}},
  \bibinfo{pages}{3964} (\bibinfo{year}{1985}),
  \urlprefix\url{http://link.aps.org/doi/10.1103/PhysRevA.31.3964}.

\bibitem[{\citenamefont{Cederbaum et~al.}(2003)\citenamefont{Cederbaum,
  Friedman, Ryaboy, and Moiseyev}}]{cederbaum}
\bibinfo{author}{\bibfnamefont{L.~S.} \bibnamefont{Cederbaum}},
  \bibinfo{author}{\bibfnamefont{R.~S.} \bibnamefont{Friedman}},
  \bibinfo{author}{\bibfnamefont{V.~M.} \bibnamefont{Ryaboy}},
  \bibnamefont{and} \bibinfo{author}{\bibfnamefont{N.}~\bibnamefont{Moiseyev}},
  \bibinfo{journal}{Phys. Rev. Lett.} \textbf{\bibinfo{volume}{90}},
  \bibinfo{pages}{013001} (\bibinfo{year}{2003}),
  \urlprefix\url{http://link.aps.org/doi/10.1103/PhysRevLett.90.013001}.

\bibitem[{\citenamefont{Schult et~al.}(1990)\citenamefont{Schult, Wyld, and
  Ravenhall}}]{schult}
\bibinfo{author}{\bibfnamefont{R.~L.} \bibnamefont{Schult}},
  \bibinfo{author}{\bibfnamefont{H.~W.} \bibnamefont{Wyld}}, \bibnamefont{and}
  \bibinfo{author}{\bibfnamefont{D.~G.} \bibnamefont{Ravenhall}},
  \bibinfo{journal}{Phys. Rev. B} \textbf{\bibinfo{volume}{41}},
  \bibinfo{pages}{12760} (\bibinfo{year}{1990}),
  \urlprefix\url{http://link.aps.org/doi/10.1103/PhysRevB.41.12760}.

\bibitem[{\citenamefont{Ji and Berggren}(1992)}]{zhen-li}
\bibinfo{author}{\bibfnamefont{Z.-L.} \bibnamefont{Ji}} \bibnamefont{and}
  \bibinfo{author}{\bibfnamefont{K.-F.} \bibnamefont{Berggren}},
  \bibinfo{journal}{Phys. Rev. B} \textbf{\bibinfo{volume}{45}},
  \bibinfo{pages}{6652} (\bibinfo{year}{1992}),
  \urlprefix\url{http://link.aps.org/doi/10.1103/PhysRevB.45.6652}.

\bibitem[{\citenamefont{Olendski and Mikhailovska}(2003)}]{olendski}
\bibinfo{author}{\bibfnamefont{O.}~\bibnamefont{Olendski}} \bibnamefont{and}
  \bibinfo{author}{\bibfnamefont{L.}~\bibnamefont{Mikhailovska}},
  \bibinfo{journal}{Phys. Rev. B} \textbf{\bibinfo{volume}{67}},
  \bibinfo{pages}{035310} (\bibinfo{year}{2003}),
  \urlprefix\url{http://link.aps.org/doi/10.1103/PhysRevB.67.035310}.

\bibitem[{\citenamefont{Rotter and Sadreev}(2005)}]{rotter}
\bibinfo{author}{\bibfnamefont{I.}~\bibnamefont{Rotter}} \bibnamefont{and}
  \bibinfo{author}{\bibfnamefont{A.~F.} \bibnamefont{Sadreev}},
  \bibinfo{journal}{Phys. Rev. E} \textbf{\bibinfo{volume}{71}},
  \bibinfo{pages}{046204} (\bibinfo{year}{2005}),
  \urlprefix\url{http://link.aps.org/doi/10.1103/PhysRevE.71.046204}.

\bibitem[{\citenamefont{Orellana and
  Dom\'{\i}nguez-Adame}(2006)}]{orellanapssa}
\bibinfo{author}{\bibfnamefont{P.}~\bibnamefont{Orellana}} \bibnamefont{and}
  \bibinfo{author}{\bibfnamefont{F.}~\bibnamefont{Dom\'{\i}nguez-Adame}},
  \bibinfo{journal}{physica status solidi (a)} \textbf{\bibinfo{volume}{203}},
  \bibinfo{pages}{1178} (\bibinfo{year}{2006}), ISSN \bibinfo{issn}{1862-6319},
  \urlprefix\url{http://dx.doi.org/10.1002/pssa.200566124}.

\bibitem[{\citenamefont{Ladr\'on~de Guevara and Orellana}(2006)}]{loreto}
\bibinfo{author}{\bibfnamefont{M.~L.} \bibnamefont{Ladr\'on~de Guevara}}
  \bibnamefont{and} \bibinfo{author}{\bibfnamefont{P.~A.}
  \bibnamefont{Orellana}}, \bibinfo{journal}{Phys. Rev. B}
  \textbf{\bibinfo{volume}{73}}, \bibinfo{pages}{205303}
  (\bibinfo{year}{2006}),
  \urlprefix\url{http://link.aps.org/doi/10.1103/PhysRevB.73.205303}.

\bibitem[{\citenamefont{Gonz\'alez
  et~al.}(2010{\natexlab{a}})\citenamefont{Gonz\'alez, Pacheco, Rosales, and
  Orellana}}]{epljhon}
\bibinfo{author}{\bibfnamefont{J.~W.} \bibnamefont{Gonz\'alez}},
  \bibinfo{author}{\bibfnamefont{M.}~\bibnamefont{Pacheco}},
  \bibinfo{author}{\bibfnamefont{L.}~\bibnamefont{Rosales}}, \bibnamefont{and}
  \bibinfo{author}{\bibfnamefont{P.~A.} \bibnamefont{Orellana}},
  \bibinfo{journal}{EPL (Europhysics Letters)} \textbf{\bibinfo{volume}{91}},
  \bibinfo{pages}{66001} (\bibinfo{year}{2010}{\natexlab{a}}),
  \urlprefix\url{http://stacks.iop.org/0295-5075/91/i=6/a=66001}.

\bibitem[{\citenamefont{Texier}(2002)}]{texier}
\bibinfo{author}{\bibfnamefont{C.}~\bibnamefont{Texier}},
  \bibinfo{journal}{Journal of Physics A: Mathematical and General}
  \textbf{\bibinfo{volume}{35}}, \bibinfo{pages}{3389} (\bibinfo{year}{2002}),
  \urlprefix\url{http://stacks.iop.org/0305-4470/35/i=15/a=303}.

\bibitem[{\citenamefont{Miyamoto}(2005)}]{Miyamoto}
\bibinfo{author}{\bibfnamefont{M.}~\bibnamefont{Miyamoto}},
  \bibinfo{journal}{Phys. Rev. A} \textbf{\bibinfo{volume}{72}},
  \bibinfo{pages}{063405} (\bibinfo{year}{2005}),
  \urlprefix\url{http://link.aps.org/doi/10.1103/PhysRevA.72.063405}.

\bibitem[{\citenamefont{Bulgakov et~al.}(2007)\citenamefont{Bulgakov, Rotter,
  and Sadreev}}]{Sadreev1}
\bibinfo{author}{\bibfnamefont{E.~N.} \bibnamefont{Bulgakov}},
  \bibinfo{author}{\bibfnamefont{I.}~\bibnamefont{Rotter}}, \bibnamefont{and}
  \bibinfo{author}{\bibfnamefont{A.~F.} \bibnamefont{Sadreev}},
  \bibinfo{journal}{Phys. Rev. A} \textbf{\bibinfo{volume}{75}},
  \bibinfo{pages}{067401} (\bibinfo{year}{2007}),
  \urlprefix\url{http://link.aps.org/doi/10.1103/PhysRevA.75.067401}.

\bibitem[{\citenamefont{Sadreev et~al.}(2006)\citenamefont{Sadreev, Bulgakov,
  and Rotter}}]{Sadreev2}
\bibinfo{author}{\bibfnamefont{A.~F.} \bibnamefont{Sadreev}},
  \bibinfo{author}{\bibfnamefont{E.~N.} \bibnamefont{Bulgakov}},
  \bibnamefont{and} \bibinfo{author}{\bibfnamefont{I.}~\bibnamefont{Rotter}},
  \bibinfo{journal}{Phys. Rev. B} \textbf{\bibinfo{volume}{73}},
  \bibinfo{pages}{235342} (\bibinfo{year}{2006}),
  \urlprefix\url{http://link.aps.org/doi/10.1103/PhysRevB.73.235342}.

\bibitem[{\citenamefont{Bulgakov and Sadreev}(2010)}]{Sadreev3}
\bibinfo{author}{\bibfnamefont{E.~N.} \bibnamefont{Bulgakov}} \bibnamefont{and}
  \bibinfo{author}{\bibfnamefont{A.~F.} \bibnamefont{Sadreev}},
  \bibinfo{journal}{Phys. Rev. B} \textbf{\bibinfo{volume}{81}},
  \bibinfo{pages}{115128} (\bibinfo{year}{2010}),
  \urlprefix\url{http://link.aps.org/doi/10.1103/PhysRevB.81.115128}.

\bibitem[{\citenamefont{N\"ockel}(1992)}]{nockel}
\bibinfo{author}{\bibfnamefont{J.~U.} \bibnamefont{N\"ockel}},
  \bibinfo{journal}{Phys. Rev. B} \textbf{\bibinfo{volume}{46}},
  \bibinfo{pages}{15348} (\bibinfo{year}{1992}),
  \urlprefix\url{http://link.aps.org/doi/10.1103/PhysRevB.46.15348}.

\bibitem[{\citenamefont{Marinica et~al.}(2008)\citenamefont{Marinica, Borisov,
  and Shabanov}}]{marinica}
\bibinfo{author}{\bibfnamefont{D.~C.} \bibnamefont{Marinica}},
  \bibinfo{author}{\bibfnamefont{A.~G.} \bibnamefont{Borisov}},
  \bibnamefont{and} \bibinfo{author}{\bibfnamefont{S.~V.}
  \bibnamefont{Shabanov}}, \bibinfo{journal}{Phys. Rev. Lett.}
  \textbf{\bibinfo{volume}{100}}, \bibinfo{pages}{183902}
  (\bibinfo{year}{2008}),
  \urlprefix\url{http://link.aps.org/doi/10.1103/PhysRevLett.100.183902}.

\bibitem[{\citenamefont{Bulgakov and Sadreev}(2008)}]{evgeny}
\bibinfo{author}{\bibfnamefont{E.~N.} \bibnamefont{Bulgakov}} \bibnamefont{and}
  \bibinfo{author}{\bibfnamefont{A.~F.} \bibnamefont{Sadreev}},
  \bibinfo{journal}{Phys. Rev. B} \textbf{\bibinfo{volume}{78}},
  \bibinfo{pages}{075105} (\bibinfo{year}{2008}),
  \urlprefix\url{http://link.aps.org/doi/10.1103/PhysRevB.78.075105}.

\bibitem[{\citenamefont{Prodanovi\'c et~al.}(2009)\citenamefont{Prodanovi\'c,
  Milanovi\'c, and Radovanovi\'c}}]{prodanovic}
\bibinfo{author}{\bibfnamefont{N.}~\bibnamefont{Prodanovi\'c}},
  \bibinfo{author}{\bibfnamefont{V.}~\bibnamefont{Milanovi\'c}},
  \bibnamefont{and}
  \bibinfo{author}{\bibfnamefont{J.}~\bibnamefont{Radovanovi\'c}},
  \bibinfo{journal}{Journal of Physics A: Mathematical and Theoretical}
  \textbf{\bibinfo{volume}{42}}, \bibinfo{pages}{415304}
  (\bibinfo{year}{2009}),
  \urlprefix\url{http://stacks.iop.org/1751-8121/42/i=41/a=415304}.

\bibitem[{\citenamefont{Gonz\'alez-Santander
  et~al.}(2013)\citenamefont{Gonz\'alez-Santander, Orellana, and
  Dom\'{\i}nguez-Adame}}]{eplclara2}
\bibinfo{author}{\bibfnamefont{C.}~\bibnamefont{Gonz\'alez-Santander}},
  \bibinfo{author}{\bibfnamefont{P.~A.} \bibnamefont{Orellana}},
  \bibnamefont{and}
  \bibinfo{author}{\bibfnamefont{F.}~\bibnamefont{Dom\'{\i}nguez-Adame}},
  \bibinfo{journal}{EPL (Europhysics Letters)} \textbf{\bibinfo{volume}{102}},
  \bibinfo{pages}{17012} (\bibinfo{year}{2013}),
  \urlprefix\url{http://stacks.iop.org/0295-5075/102/i=1/a=17012}.

\bibitem[{\citenamefont{Plotnik et~al.}(2011)\citenamefont{Plotnik, Peleg,
  Dreisow, Heinrich, Nolte, Szameit, and Segev}}]{plotnik}
\bibinfo{author}{\bibfnamefont{Y.}~\bibnamefont{Plotnik}},
  \bibinfo{author}{\bibfnamefont{O.}~\bibnamefont{Peleg}},
  \bibinfo{author}{\bibfnamefont{F.}~\bibnamefont{Dreisow}},
  \bibinfo{author}{\bibfnamefont{M.}~\bibnamefont{Heinrich}},
  \bibinfo{author}{\bibfnamefont{S.}~\bibnamefont{Nolte}},
  \bibinfo{author}{\bibfnamefont{A.}~\bibnamefont{Szameit}}, \bibnamefont{and}
  \bibinfo{author}{\bibfnamefont{M.}~\bibnamefont{Segev}},
  \bibinfo{journal}{Phys. Rev. Lett.} \textbf{\bibinfo{volume}{107}},
  \bibinfo{pages}{183901} (\bibinfo{year}{2011}),
  \urlprefix\url{http://link.aps.org/doi/10.1103/PhysRevLett.107.183901}.

\bibitem[{\citenamefont{Capasso et~al.}(1992)\citenamefont{Capasso, Sirtori,
  Faist, Sivico, Chu, and Cho}}]{capasso}
\bibinfo{author}{\bibfnamefont{F.}~\bibnamefont{Capasso}},
  \bibinfo{author}{\bibfnamefont{C.}~\bibnamefont{Sirtori}},
  \bibinfo{author}{\bibfnamefont{J.}~\bibnamefont{Faist}},
  \bibinfo{author}{\bibfnamefont{D.~L.} \bibnamefont{Sivico}},
  \bibinfo{author}{\bibfnamefont{S.-N.~G.} \bibnamefont{Chu}},
  \bibnamefont{and} \bibinfo{author}{\bibfnamefont{A.~Y.} \bibnamefont{Cho}},
  \bibinfo{journal}{Nature} \textbf{\bibinfo{volume}{358}},
  \bibinfo{pages}{565} (\bibinfo{year}{1992}).

\bibitem[{\citenamefont{de~Longhi and G.}(2013)}]{repdelonghi}
\bibinfo{author}{\bibfnamefont{S.}~\bibnamefont{de~Longhi}} \bibnamefont{and}
  \bibinfo{author}{\bibfnamefont{D.~V.} \bibnamefont{G.}},
  \bibinfo{journal}{Sci. Rep.} \textbf{\bibinfo{volume}{3}},
  \bibinfo{pages}{02219} (\bibinfo{year}{2013}).

\bibitem[{\citenamefont{Zhang et~al.}(2012)\citenamefont{Zhang, Braak, and
  Kollar}}]{zhangprl}
\bibinfo{author}{\bibfnamefont{J.~M.} \bibnamefont{Zhang}},
  \bibinfo{author}{\bibfnamefont{D.}~\bibnamefont{Braak}}, \bibnamefont{and}
  \bibinfo{author}{\bibfnamefont{M.}~\bibnamefont{Kollar}},
  \bibinfo{journal}{Phys. Rev. Lett.} \textbf{\bibinfo{volume}{109}},
  \bibinfo{pages}{116405} (\bibinfo{year}{2012}),
  \urlprefix\url{http://link.aps.org/doi/10.1103/PhysRevLett.109.116405}.

\bibitem[{\citenamefont{Liu et~al.}(2009)\citenamefont{Liu, Suenaga, Harris,
  and Iijima}}]{Liu_2009}
\bibinfo{author}{\bibfnamefont{Z.}~\bibnamefont{Liu}},
  \bibinfo{author}{\bibfnamefont{K.}~\bibnamefont{Suenaga}},
  \bibinfo{author}{\bibfnamefont{P.~J.~F.} \bibnamefont{Harris}},
  \bibnamefont{and} \bibinfo{author}{\bibfnamefont{S.}~\bibnamefont{Iijima}},
  \bibinfo{journal}{Phys. Rev. Lett.} \textbf{\bibinfo{volume}{102}},
  \bibinfo{eid}{015501} (\bibinfo{year}{2009}).

\bibitem[{\citenamefont{Orellana et~al.}(2013)\citenamefont{Orellana, Rosales,
  Chico, and Pacheco}}]{JAP2013}
\bibinfo{author}{\bibfnamefont{P.~A.} \bibnamefont{Orellana}},
  \bibinfo{author}{\bibfnamefont{L.}~\bibnamefont{Rosales}},
  \bibinfo{author}{\bibfnamefont{L.}~\bibnamefont{Chico}}, \bibnamefont{and}
  \bibinfo{author}{\bibfnamefont{M.}~\bibnamefont{Pacheco}},
  \bibinfo{journal}{Journal of Applied Physics} \textbf{\bibinfo{volume}{113}},
  \bibinfo{eid}{213710} (\bibinfo{year}{2013}),
  \urlprefix\url{http://scitation.aip.org/content/aip/journal/jap/113/21/10.1063/1.4809752}.

\bibitem[{\citenamefont{Gonz\'alez
  et~al.}(2010{\natexlab{b}})\citenamefont{Gonz\'alez, Santos, Pacheco, Chico,
  and Brey}}]{Jhon_2010}
\bibinfo{author}{\bibfnamefont{J.~W.} \bibnamefont{Gonz\'alez}},
  \bibinfo{author}{\bibfnamefont{H.}~\bibnamefont{Santos}},
  \bibinfo{author}{\bibfnamefont{M.}~\bibnamefont{Pacheco}},
  \bibinfo{author}{\bibfnamefont{L.}~\bibnamefont{Chico}}, \bibnamefont{and}
  \bibinfo{author}{\bibfnamefont{L.}~\bibnamefont{Brey}},
  \bibinfo{journal}{Phys. Rev. B} \textbf{\bibinfo{volume}{81}},
  \bibinfo{pages}{195406} (\bibinfo{year}{2010}{\natexlab{b}}).

\bibitem[{\citenamefont{Gonz\'alez et~al.}(2012)\citenamefont{Gonz\'alez,
  Pacheco, Orellana, Brey, and Chico}}]{Jhon_2012}
\bibinfo{author}{\bibfnamefont{J.~W.} \bibnamefont{Gonz\'alez}},
  \bibinfo{author}{\bibfnamefont{M.}~\bibnamefont{Pacheco}},
  \bibinfo{author}{\bibfnamefont{P.}~\bibnamefont{Orellana}},
  \bibinfo{author}{\bibfnamefont{L.}~\bibnamefont{Brey}}, \bibnamefont{and}
  \bibinfo{author}{\bibfnamefont{L.}~\bibnamefont{Chico}},
  \bibinfo{journal}{Solid State Communications} \textbf{\bibinfo{volume}{152}},
  \bibinfo{pages}{1400} (\bibinfo{year}{2012}).

\bibitem[{\citenamefont{Chico et~al.}(2012)\citenamefont{Chico, Gonz\'alez,
  Santos, Pacheco, and Brey}}]{Acta_2012}
\bibinfo{author}{\bibfnamefont{L.}~\bibnamefont{Chico}},
  \bibinfo{author}{\bibfnamefont{J.~W.} \bibnamefont{Gonz\'alez}},
  \bibinfo{author}{\bibfnamefont{H.}~\bibnamefont{Santos}},
  \bibinfo{author}{\bibfnamefont{M.}~\bibnamefont{Pacheco}}, \bibnamefont{and}
  \bibinfo{author}{\bibfnamefont{L.}~\bibnamefont{Brey}},
  \bibinfo{journal}{Acta Physica Polonica A} \textbf{\bibinfo{volume}{122}},
  \bibinfo{pages}{299} (\bibinfo{year}{2012}).

\end{thebibliography}

\end{document}